\documentclass[a4paper,11pt,onecolumn]{article}
\usepackage{latexsym}
\usepackage{amssymb, amsmath}
\usepackage{graphicx}
\usepackage{dcolumn}
\usepackage{bm}
\def\be{\begin{equation}}
\def\ee{\end{equation}}
\def\beq{\begin{eqnarray}}
\def\eeq{\end{eqnarray}}
\def\bn{\begin{eqnarray*}}
\def\en{\end{eqnarray*}}

\def\P{\Phi}
\def\p{\phi}
\def\w{\omega}

\def\a{\alpha}
\def\b{\beta}
\def\s{\sigma}

\def\d{\delta}

\def\g{\gamma}

\def\k{\kappa}

\def\si{\psi}

\def\e{\epsilon}
\def\n{\nu}
\def\m{\mu}
\def\r{\rho}

\def\l{\lambda}
\def\L{\Lambda}

\def\draftversion{N}                
\def\note[#1]#2{\message{(#1)}\if\draftversion
{\noindent\em[#2]\/}\fi}
\if \draftversion Y
    \fi 
\begin{document}
\title{ HOLST ACTIONS FOR  SUPERGRAVITY THEORIES}
\author{
Romesh K. Kaul \\
The Institute of Mathematical Sciences,\\
Chennai 600 113, India.\\
kaul@imsc.res.in
}

\maketitle

\begin{abstract} 
Holst action containing Immirzi parameter for pure gravity is 
generalised to the supergravity theories.  Supergravity 
equations of motion are not modified by such generalisations, 
thus preserving supersymmetry.  Dependence on the 
Immirzi parameter does not emerge in the classical equations of
motion. This is in contrast with the recent observation of 
Perez and Rovelli for gravity action containing  original Holst 
term and a  minimally coupled Dirac fermion  where the classical 
equations of motion do develop a dependence on Immirzi parameter.

\end{abstract}

\section{\bf Introduction}

In the first order formalism, pure gravity is described through
three coupling constants;  while two of them, the Newton's 
gravitational  
and the cosmological constants, are dimensionful;  the third known
as Immirzi parameter  is dimensionless. In the action these are 
associated with the Hilbert-Palatini, cosmological  and Holst
terms  respectively. Ignoring the cosmological term, 
we present the  Holst's generalisation \cite{holst} of 
the Hilbert-Palatini action in the natural system of fundamental
units where Newton's constant
$G =1/(8\pi)$ as{\footnote {Our conventions are: 
Latin indices in the beginning of alphabet,  $ a,b,c, ...$,  run
over $ 1, 2, 3, 4$ and  $e^a_\m e^{b\m}
= \d^{ab}$, $ e_\m^a e_\n^a = g_{\m\n}$. The tetrad component $e^4_\m$ is 
imaginary so are 
the connection components $\w_\m^{4 i} ~(i = 1,2, 3) $
and the determinant $e$ of  tetrad $e^a_\m$, ~$e^*= -e$
$= -{\frac 1 {4!}} ~\e^{\m\n\a\b}_{} \e_{abcd}^{} e^a_\m e^b_\n e^c_\a e^d_\b$.
The usual antisymmetric Levi-Civita density of weight one
$\e^{\m\n\a\b}_{}$ has values $\pm~1$ or $0$ and
$\e_{\m\n\a\b}^{}$ takes values $\pm ~e^2$ or $0$; and
completely antisymmetric ~
$\e^{abcd}_{} =\e_{abcd}^{}$ are $\pm~1$ or $0$.}}: 

\be
S ~ = ~ {\frac 1 2} \int d^4x ~e ~ \Sigma^{\m \n}_{ a b} 
\left[ R_{\m\n}^{~~~a b}(\w) ~ +~ i\eta ~\tilde{R}_{\m\n}^{~~~a b}(\w) 
\right]       \label{holst}
\ee
where $ \Sigma ^{ab}_{\m\n} ~=~ {\frac 1 2}~ e^a_{[\m}~ e^b_{\n]}$ and
$R_{\m\n}^{~~~ab}(\w)~=~\partial_{[\m} \w_{\n]}^{~~ab}~ +~ \w_{[\m}^{~~ac}~
\w_{\n]}^{~~cb}$. The second term containing the parameter $\eta$ is the Holst action
with $\tilde{R}_{\m\n}^{~~~a b} = {\frac 1 2} \e^{a b c d} 
R_{\m \n cd}$ and $\eta^{-1}$  is the Immirzi parameter \cite{imm}. 
For $\eta= -i$, the action (\ref{holst}) leads to  the self-dual Ashtekar 
canonical formalism for gravity in terms of complex $SU(2)$ 
connection \cite{ash}. 
For real $\eta$,  this  action allows 
a Hamiltonian formulation  \cite{holst, barros}
in terms of real $SU(2)$ connection which coincides with  
that of  Barbero \cite{barbero} for $\eta =1$.

In the first order formalism,  equations of motion are obtained 
by varying the Hilbert-Palatini-Holst action (\ref{holst})
with respect to the connection $\w_{\m}^{~ab}$ and tetrad 
$ e_\m^a$ fields independently. Variation with
respect to $\w_\m^{~ab}$ leads to the standard no-torsion equation:
$D_{[\m}^{} (\w) ~e_{\n]}^a ~ $ $=~ 0$, 
which can be solved for the connection  in terms of
tetrad fields in the usual way: $\w ~=~ \w(e)$ where the
standard spin connection is:
\be
\w_\m^{~ab} (e)~=~ {\frac 1 2} \left[ e^{\n a}\partial_{[\m}e_{\n]}^b ~-~e^{\n b}
\partial_{[\m}e_{\n]}^a ~-~e^{\r a}e^{\s b}\partial_{[\r}e_{\s]}^c e_\m^c\right]
  \label{spine}
\ee

Variation of action (\ref{holst}) with respect to the tetrad $e_\m^a$
leads to the usual Einstein equation:
$R^{~~\m}_{a} ~ $ $ -~ {\frac 1 2} ~e^\m_a ~ R $ $ ~=~0 $.
Thus, adding the Holst action to  Hilbert-Palatini action as in 
Eqn.(\ref{holst}) does not
change the equations of motion of the theory.
Notice that for $\w ~  = ~ \w(e)$,  the Holst term in the Lagrangian
density is identically zero:
~$e~ \Sigma^{\m\n}_{a b}~ {\tilde R}_{\m\n}^{~~~ab} (\w(e))$ $~=~
 $ $=~ {\frac 1 2}~
\e^{\m\n\a\b} R_{\m\n\a\b}(\w(e)) $~$=~ 0$,
due to the cyclicity property $ R_{[\m \n \a ] \b} (\w(e)) $ $ ~=~0 $.

While classical equations of motion do not depend on the Immirzi
parameter, non-perturbative physical effects depending on this 
parameter  are expected to appear in quantum gravity.

Inclusion of spin $1/2$ fermions into the Holst's generalised Hilbert-Palatini
action (\ref{holst}) has been done recently by Perez and Rovelli
and also by Freidel, Minic and Takeuchi\cite {perez-rovelli}. This has 
been achieved by  minimal coupling of  
the fermion through  a term ~~~$ - ~ (1/2) ~( {\bar \l}~\g^\m D_\m(\w) \l 
$ $~-~$  $ {\overline {D_\m^{}(\w)  \l}}~ \g^\m_{} \l) $
into the  action (\ref{holst})
without changing the Holst term. This indeed does  change 
equations of motion leading to dependence on the Immirzi parameter 
even at classical level.
However, as shown by Mercuri \cite{mercuri}, it is possible to  
modify the Holst action in presence
of Dirac fermions so that the classical equations motion stay independent
of the Immirzi parameter. To do this, to the Einstein-Cartan
action{\footnote{ In our conventions all the 
Dirac gamma matrices are
hermitian $(\g^a)^\dagger = \g^a$, ~$\g^a \g^b + \g^b \g^a = 2\d ^{ab}$
and ~$\g_5 = \g^1 \g^2 \g^3 \g^4$,
~$(\g_5)^2 = +1$ and $\s_{ab} = {\frac 1 2} \g_{[a} \g_{b]}$.
For Majorana fermions ${\bar \si} = \si^T C$
where $C$ is the charge conjugation matrix with properties $ C^\dagger C
= C C^\dagger = 1$, ~$ C^T = - C$, ~ $C\g_a C^{-1} = -\g_a^T$.}}:
\be
S_{GF}^{~} ~=~ ~ {\frac 1 2} \int d^4x ~e ~ \left [\Sigma^{\m \n}_{ a b}
 R_{\m\n}^{~~~a b}(\w)  -  {\bar \l}\g^\m D_\m(\w) \l
+ {\overline {D_\m(\w) \l}} ~\g^\m \l \right], \label{GF}
\ee
\noindent we add a modified Holst term introducing
a non-minimal coupling  for the fermion:
\be
S_{HolstF}^{~}~=~ {\frac {i\eta} 2} \int d^4x ~e  \left[ \Sigma^{\m \n}_{ a b}
{\tilde R}_{\m\n}^{~~~a b}(\w)  -  {\bar \l} \g^{}_5 
\g^\m D_\m(\w) \l
- {\overline {D_\m(\w) \l}}~ \g^{~}_5 \g^\m \l \right] \label{HolstF}
\ee

Variation of the total action $ S_{GF}^{~} + S_{HolstF}^{~}$ with
respect to the connection field $\w_\m^{~ab}$ 
yields the standard torsion equation as an equation of motion:

\beq
D^{}_{[\m} (\w) e^a_{\n ]} ~=~ 2~T_{\m\n}^{~~~a}(\l)~ \equiv ~{\frac 1 {2e}} 
e^{ a \a}_{~} ~\e_{\m\n\a\b}^{}~ {\bar \l}\g_5^{~} \g^\b_{~} \l
\label{torsionF}
\eeq
\noindent This can be solved as:

\be
\w^{~}_{\m ab} ~=~ \w_{\m ab}^{}(e, \l) \equiv \w^{~}_{\m ab}(e) ~+~ \k^{~}_{\m ab}(\l) \label{spinF}
\ee
\noindent where $\w(e)$ the spin connection of pure gravity (\ref{spine}) 
and the contorsion tensor  is given by
(general relation between torsion and contorsion is
$2~T_{\m\n}^{~~~\l} = - \k_{[\m\n]}^{~~~~\l}$):
\be
\k^{~}_{\m ab}(\l) ~=~ -~ {\frac 1 4}~e^c_\m~ \e^{~}_{abcd} ~{\bar \l} 
\g^{~}_5 \g^d_{~} \l \label{contorsionF}
\ee

It is straight forward to check that the fermionic Holst Lagrangian
density (\ref{HolstF}) above is a total derivative for connection
$\w(e, \l) = \w(e) + \k(\l)$ given by (\ref{spinF}) and (\ref{contorsionF}).
Mercuri has made an interesting observation\cite{mercuri} that the modified
Holst action $S_{HolstF}^{~}[\w(e, \l)]$ can be cast in a 
form involving
the Nieh-Yan invariant density and divergence of an axial current
density in the following manner:
\beq
S^{~}_{HolstF} \left[ \w(e, \l) \right] = ~-~{\frac {i\eta} 2}
\int d^4x  \left[ I_{NY}^{} ~  +~ \partial_\m^{~}  J^\m_{~}(\l) \right] \label{mercuriF}
\eeq
\noindent where $J_\m^{} (\l) = e~{\bar \l}\g_5^{~}\g^{~}_\m \l$ 
 and the Nieh-Yan invariant density in general is 
\cite{nieh-yan}:
\be
I^{}_{NY}~=~ \e^{\m\n\a\b} \left[ ~ T_{\m\n}^{~~a}~T_{\a\b a}~-~
{\frac 1 2}~\Sigma^{ab}_{\m\n}
~R^{~}_{\a\b ab}(\w)~ \right]
\label{nieh-yan}
\ee

For the present case, notice that $\e^{\m\n\a\b}_{~} T_{\m\n }^{~~a}(\l)
T_{\a\b a}(\l)$ is identically zero for the explicit torsion expression
of Eqn.(\ref{torsionF}) and hence Nieh-Yan invariant density is simply
$-~(1/2)~\e^{\m\n\a\b}_{}~ \Sigma_{\m\n}^{ab}~ R^{}_{\a\b ab} (\w(e, \l))$.
In general the Nieh-Yan topological invariant density is just 
 the divergence of  pseudo-trace axial vector
constructed from torsion:
\be
 I^{}_{NY} ~=~ \e^{\m\n\a\b}_{}~ ~\partial_\m^{} T^{}_{\n\a\b} 
 \label{nieh-yan1}
\ee
\noindent This allows us to see that the modified Holst Lagrangian
density is
indeed a total derivative when the connection equation of motion
(\ref{spinF} and \ref{contorsionF}) is used:
\bn
S^{~}_{HolstF} \left[ \w(e, \l) \right] = {\frac {i\eta} 4}
\int d^4x ~\partial^{~}_\m  J^\m_{} (\l) =
- {\frac {i\eta} 6} \int d^4x ~\e^{\m\n\a\b}_{} \partial_\m^{}
T_{\n\a\b}^{}(\l)
\en
\noindent where we have used the fact that $2~\e^{\m\n\a\b}_{}~
T_{\n\a\b}^{}(\l)$ $ = -3~ J^\m_{}(\l)$.

 Next variations of the total action
$ S_{GF}^{~} + S_{HolstF}^{~}$ with respect to tetrad field 
$e^a_\m$ and  fermion $\l$  lead 
to same equations of motion as those obtained from the variations of
gravity-fermion action $S_{GF}^{~}$ alone, making
these classical equations of motion independent of  Immirzi
parameter.

Coupling of higher spin fermions to gravity also requires a special
consideration in presence of Holst term. For example, we could
consider the supergravity theories which contain spin $3/2$ fermions.
If we add the original Holst term of Eqn.(\ref{holst}) without 
any modifications to the standard actions of these theories
in the manner done by Perez and Rovelli \cite{perez-rovelli} for spin $1/2$ 
fermions,  the equations of motion obtained from the resulting actions
will indeed develop dependence
on Immirzi parameter indicating violation of supersymmetry.
It is worthwhile to ask if there are any  possible modifications
of the Holst term which preserve the original supergravity
equations of motion.  In the following we shall discuss such 
modifications of Holst action which when added to the standard
$N =1, 2,4$ supergravity actions will  leave 
supergravity equations of motion unchanged and thereby preserve 
supersymmetry. In addition we shall also see that in each of these
cases, for the  connection satisfying  connection equation of motion
the modified Holst action can be written in an analogous 
form as written by Mercuri for spin $ 1/2$ fermions (\ref{mercuriF}).

\section{N~=~1  supergravity with Holst action}

The simplest supersymmetric generalisation of Einstein gravity
is $N = 1$ supergravity \cite{N1SUGRA} which is described by a
spin ${\frac 3 2}$ Majorana spinor, the gravitino $\si_\m$,
and the tetrad field $e^a_\mu$.
Generalised  supergravity action containing the modified 
Holst term for this theory is given by:
\be
 S_1 ~ =~ S_{SG1} ~ +~ S_{SHolst1}  \label{N1action}
\ee
where the supergravity action is:

\be
S_{SG1} ~=~ {\frac 1 2} \int d^4x  ~ \left[~ e~\Sigma^{\m \n}_{ a b}
~R_{\m\n}^{~~~a b}(\w) ~ -~ 
\epsilon^{\mu \nu \alpha \beta} ~{\bar \psi}_\mu \gamma_5
\gamma_\nu D_\alpha ( \w) \psi_\beta \right] \label{N1SGaction}
\ee
\noindent and supersymmetric Holst action as introduced by Tsuda 
\cite{N1sholst} is:
\be
S_{SHolst1} ~ = ~ {\frac {i\eta} 2} \int d^4x~\left[~ e~\Sigma^{\m \n}_{ a b}
~{\tilde R}_{\m\n}^{~~~a b}(\w) ~ -~
\epsilon^{\mu \nu \alpha \beta} ~{\bar \psi}_\mu 
\gamma_\nu D_\alpha ( \w) \psi_\beta \right]   \label{N1Haction}
\ee
\noindent Again for $\eta~ =~ - i$, action (\ref{N1action}) is the
$N=1$ supersymmetric generalisation of the Ashtekar chiral action.

Variation of action $ S_1$  with respect to 
connection $\w_\m^{~ab}$ leads to
the standard torsion equation of $N=1$ supergravity:
\be
D_{[\m}^{}(\w)~e^a_{\n]} ~ =~2~ T_{\m\n}^{~~~a}(\si) ~\equiv ~
{\frac 1 2}~{\bar \si}_\m \g^a \si_\n \label{N1torsion}
\ee
\noindent which in turn is solved by
\be
\w_\m^{~ab} ~ =~ \w_\m^{~ab} (e, \si) \equiv \w_\m^{~ab} (e) ~ 
+~ \k_\m^{~ab}(\si)
\ee
where $\w(e)$ is the pure gravity spin-connection given by (\ref{spine}) and the contorsion tensor is
\be
\k_{\m\a\b}(\si)~=~ {\frac 1 4}~\left[ ~ {\bar \si}_\a \g_\m \si_\b 
~+~{\bar \si}_\m\g_\a \si_\b ~-~ {\bar \si}_\m \g_\b \si_\a ~\right]
\ee

Next, supersymmetric Holst Lagrangian density(\ref{N1Haction}) 
is a total derivative for $\w= \w (e, \si) $. It can also be cast
in the form as in (\ref{mercuriF}) involving  Nieh-Yan topological 
invariant density and divergence of an axial current density as:
\beq
S^{~}_{SHolst1} \left[  \w(e, \si) \right]~ =~-~ {\frac {i\eta} 2}
\int d^4x \left[ I^{}_{NY}~+~
\partial_\m^{~}  J^\m_{~} (\si) \right] 
\label{N1nieh-yan}\eeq
\noindent where now we have the gravitino  axial vector current density 
$J^\m_{}(\si) = {\frac  1 2}~ \e^{\m\n\a\b}_{} {\bar \si}_\n^{}\g_\a^{}\si_\b^{} $.
\noindent Here also, Fierz rearrangement implies
$\e^{\m\n\a\b}_{~} T_{\m\n a}^{}(\si) T_{\a\b }^{~~~a}(\si)$  $=0$ 
for the torsion given by (\ref{N1torsion}) and hence
the Nieh-Yan density is simply $ - ~(1/2)~\e^{\m\n\a\b}_{}~
\Sigma^{ab}_{\a\b}~
R^{}_{\m\n ab}(\w(e,\si))$. Using the general property 
of the Nieh-Yan topological
invariant density given in Eqn.(\ref{nieh-yan1}), it follows that
the modified Holst Lagrangian density for the connection $\w(e, \si)$
is a total derivative:
\bn
S_{SHolst1} [\w(e,\si) ]= ~-~{\frac {i\eta} 4}\int d^4x 
~\partial_\m J^\m_{}(\si) =  {\frac {i\eta} 2} \int d^4x~ 
\e^{\m\n\a\b}_{} \partial_\m^{} T_{\n\a\b}^{}(\si)
\en
\noindent This is to be contrasted with the pure gravity case
above where Holst Lagrangian density is exactly zero for $\w = \w(e)$.

When substitution $ \w= \w(e,\si)$ is made into the variation of  
super-Holst action (\ref{N1Haction}) with respect to
gravitino $\si_\m$ and tetrad $e_\m^a$ fields, we obtain integrals
over total derivatives and hence these do not contribute to the equations
of motion which come entirely from the variations of  supergravity 
action $S_{SG1}$ (\ref{N1SGaction}).
Thus  addition of  super-Holst action (\ref{N1Haction})
to supergravity action (\ref{N1SGaction}) does not change the standard equations of motion of
$N=1$ supergravity.

\section{N~=~2 super-Holst action}

Next level supersymmetric generalisation of Einstein gravity is
the $N=2$ supergravity \cite{N2SUGRA}. Besides  the tetrad fields
$e^a_\m$ and their two super-partner gravitinos
whose chiral projections are $\si_\m^I$ and $\si^{~}_{I\m}$,  $I =1,2$
~($\g_5 \si^I_\m 
= + \si^I_\m$ and $ \g_5 \si^{~}_{I\m} = - \si^{~}_{I\m}$), this theory
also contains an Abelian gauge field $ A_\m$.
The action for this theory is given by \cite{N2SUGRA}:

\beq
S^{~}_{SG2} & =&  \int d^4x ~e~ \left[ \right.  {\frac 1 2} ~\Sigma^{\m\n}_{ab} 
~R_{\m\n}^{~~~ab} (\w) ~ -~ {\frac  1 4} ~ F_{\m\n}~  F^{\m\n} \cr
&&~~~~~~~~~~~~~~-~{\frac 1 {2e}}~ \e^{\m\n\a\b}\left({\bar \si}^I_\m \g^{~}_\n D_\a^{~}
 (\omega) \si^{~}_{I\b} - {\bar \si}^{~}_{I\m} \g^{~}_\n D_\a^{~}
 (\omega) \si_\b^I \right)  \cr  
&&~~~~~~~~~~~~~~+~ {\frac 1 {2\sqrt2}}~ {\bar \si}^I_\m \si^J_\n \e^{}_{IJ}
 (F^{+\m\n} + {\hat F}^{+\m\n}) \cr
&&~~~~~~~~~~~~~~+~ \left. {\frac 1 {2\sqrt2}}~ {\bar \si}^{~}_{I\m} \si^{~}_{J\n} \e^{IJ}  (F^{-\m\n}
 + {\hat F}^{-\m\n} ) \right]  \label{N2SGaction}
\eeq

\noindent where super-covariant field strength is
\bn
{\hat F}_{\m\n} ~=~ \partial_{[\m} A_{\n]} ~ -~ {\frac 1 {\sqrt 2}}~
\left( {\bar \si}_\m^I \si^J_\n \e^{~}_{IJ} ~+~ {\bar\si}^{~}_{I\m} \si^{~}_{J\n} \e^{IJ} \right)
\en
\noindent and self-(antiself-)dual field strengths are $ F_{\m\n}^\pm 
= {\frac 1 2} (F_{\m\n} \pm  ^*F_{\m\n})$ and star dual $^*$ is
given by ~$^*F_{\m\n} = {\frac 1 {2e}}~ \e_{\m\n\a \b} ~F^{\a\b}$.

We generalise the $N=2$ supergravity action (\ref{N2SGaction}) by adding a
modified Holst term to obtain the new action as:

\be
S_2~=~ S^{~}_{SG2} ~+~ S^{~}_{SHolst2}  \label{N2action}
\ee
\noindent where the super-Holst action is
\beq
S^{~}_{SHolst2} ~=~ i\eta \int d^4x ~e~ \left[ {\frac 1 2}~ \Sigma^{\m\n}_{a b}
~{\tilde R}_{\m\n}^{~~~ab}(\w)
~-~{\frac 1 {4e}} ~\e^{\m\n\a\b} {\bar \si}^I_\m \si^J_\n 
~{\bar \si}^{~}_{I\a} \si^{~}_{J\b} \right.  \cr
~ -~ \left. {\frac 1 {2e}}~e^{\m\n\a\b} \left(
{\bar \si}^I_\m \g^{~}_\n D^{~}_\a(\w)\si^{~}_{I\b}~ +~ {\bar \si}^{~}_{I\m}
\g^{~}_\n D^{~}_\a(\w)\si^I_\b \right) \right]  \label{N2Haction}
\eeq

Notice that this $N=2$ super-Holst action has an additional 
four-gravitino
term as compared to similar $N=1$ super-Holst action (\ref{N1Haction}).
This term plays an important role as shall be seen in what follows.
Also, in this modified Holst action, 
there are only fields that couple to the connection field 
$\w$ in the original supergravity action;
no terms involving the gauge field $A_\m^{}$  are included. 
This modified Holst action
as it is does have the desired property of leaving the original
supergravity equations unaltered.
To see this, we vary the generalised total action $S_2$
(\ref{N2action})
with respect to the connection $\w_\m^{~ab}$ to obtain:

\bn
-{\frac 1 2} \int d^4x ~\e^{\m\n\a\b}\left[ D^{~}_\m(\w)\Sigma^{ab}_{\a\b}
+ e^a_\m {\bar \si}^I_\a \g^b_{~}\si^{~}_{I\b} \right] \left({\frac 1 2}\e_{abcd}  + i\eta \d_{ac} \d_{bd} \right) \d \w_\n^{~cd} ~=~0
\en
\noindent which implies:
\bn
\e^{\m\n\a\b}~D^{~}_\m(\w)\Sigma^{ab}_{\a\b} ~=~ -~{\frac 1 2}~ \e^{\m\n\a\b}
e^{[a}_\m {\bar \si}^I_\a \g^{b]}_{~}\si^{~}_{I\b}
\en
which in turn leads to the standard torsion equation of $N=2$ supergravity:
\bn
D^{~}_{[\m}(\w)e^a_{\n]}~ =~ 2~ T_{\m\n}^{~~~a}(\si) \equiv
{\frac 1 2}~ \left( {\bar \si}^I_\m \g^a_{~}
\si^{~}_{I\n} ~+~ {\bar \si}^{~}_{I\m} \g^a_{~} \si^I_\n \right)
\label{N2torsion}
\en
\noindent whose solution is given by:
\be
\w_\m^{~ab} ~= ~ \w_\m^{~ab}(e, \si)~\equiv~ \w_\m^{~ab}(e) ~ +~ 
\k_\m^{~ab}(\si) \label{N2spin}
\ee
\noindent Here $\w_\m^{~ab}(e)$ is the usual torsion-free spin-connection 
(\ref{spine}) and contorsion tensor of $N=2$ supergravity is:
\be
\k_{\m\a\b}(\si) ~=~ {\frac 1 4} \left[ {\bar \si}^I_\a \g^{~}_\m \si^{~}_{I\b}
~+~ {\bar \si}^I_\m \g^{~}_\a \si^{~}_{I\b} ~-~ {\bar \si}^I_\m 
\g^{~}_\b \si^{~}_{I\a} ~ + ~ c.c. \right] \label{N2contorsion}
\ee

\noindent Thus, despite the additional super-Holst term $S^{}_{SHolst2}$
in the total action $S^{}_2$ above,
the connection equations (\ref{N2spin}, \ref{N2contorsion}) obtained
are  the standard $N=2$ supergravity equations.

Next for this connection $\w(e, \si)$, the super-Holst Lagrangian density 
 (\ref{N2Haction}) is a total derivative. To see this, notice that:
\beq
-~{\frac 1 2}~\e^{\m\n\a\b}\left[ {\bar \si}^I_\m \g^{~}_\n D^{~}_\a
(\w(e, \si)) \si^{~}_{I\b} + {\bar \si}^{~}_{I\m} \g^{~}_\n D^{~}_\a
(\w(e,\si)) \si_\b^I  + {\frac 1 2}  {\bar \si}^I_\m \si^J_\n 
{\bar \si}^{~}_{I\a}\si^{~}_{J\b} \right] ~~ \cr
~=~-~{\frac 1 2} \left[ \partial_\m^{}J^\m_{}(\si) +
 \e^{\m\n\a\b}T_{\m\n a}^{} T_{\a\b}^{~~~a}  \right]  \label{N2Haction2}
 ~~~~
\eeq
\noindent where the axial current density $J^\m_{}(\si) =  \e^{\m\n\a\b}
{\bar \si}_\n^{I}\g_\a^{}\si^{}_{I\b}$. To obtain this relation 
we have made use of $ 2~T_{\m\n}^{~~~\l} ~=~ -~\k_{[\m\n]}^{~~~\l}$ 
$= {\frac 1 2} ~{\bar \si}^I_{[\m}\g^\l_{}\si^{}_{\n ]I}$ and the  identity:
\bn
-~{\frac 1 2}~\e^{\m\n\a\b}~T_{\m\n a}^{}(\si) T_{\a\b}^{~~~a} (\si)
~=~ {\frac 1 4} ~\e^{\m\n\a\b}~
{\bar \si}^I_\m \si^J_\n {\bar \si}^{~}_{I\a}\si^{~}_{J\b} 
\en
\noindent which can be checked easily using the explicit
expression for the torsion and a simple
Fierz rearrangement. 
Clearly the four-gravitino term in the left hand side of
Eqn.(\ref{N2Haction2})
which has its origin the the four-gravitino term in the
super-Holst action (\ref{N2Haction}) is important to obtain 
the desired form of this equation.

Here also for the connection $\w(e, \si)$ given by (\ref{N2spin})
and (\ref{N2contorsion}), the super-Holst Lagrangian density can be
written in a special form in terms of the Nieh-Yan invariant 
density and divergence of an axial current density as:
\beq
S^{~}_{SHolst2} \left[  \w(e, \si)\right] ~=~ -~{\frac {i\eta} 2}
\int d^4x 
\left[ ~I_{NY}^{} ~+~ 
\partial_\m^{~}  J^\m_{~}(\si) ~\right]
\label{N2nieh-yan}\eeq
\noindent Again using the general property of the Nieh-Yan
invariant density relating it to a derivative of torsion (\ref{nieh-yan1}),
we find  that super-Holst Lagrangian density is a total 
derivative for connection $\w(e,\si)$: 
\beq
S^{~}_{SHolst2} \left[ \w(e, \si) \right] 
= -{\frac {i\eta} 4}\int d^4x~ \partial_\m^{}  J^\m_{}(\si)
=~{\frac {i\eta} 2} 
\int d^4x ~\e^{\m\n\a\b}~ \partial_\m T^{}_{\n\a\b} (\si) 
\eeq
\noindent where we have used the fact that $ 2\e^{\m\n\a\b}_{} T_{\n\a\b}^{}
(\si)$ $ = - J^\m_{}(\si)$.

Not only the connection equation 
of $N=2$ supergravity is unchanged  by adding the super-Holst 
action (\ref{N2Haction}), other equations of motion are also
not modified. For example, to check this explicitly,
substituting $\w= \w(e, \si) =\w(e) +\k(\si)$ into the variation of 
super-Holst Lagrangian density ${\cal L}_{SHolst2}$ (\ref{N2Haction})
with respect to gravitino field $\si^I_\m$ leads to:
\bn
&& \left[ \d \si^I_\m ~{\frac {\d {\cal L}^{~}_{SHolst2}} {\d \si^I_\m}}  
\right]_{\w=\w(e, \si)} ~\cr
&&~~~~=~ -~{\frac {i\eta} 2} ~\e^{\m\n\a\b} ~
\left[ \right. \d {\bar\si}^I_\m \g^{~}_\n D^{~}_\a(\w(e)) \si^{~}_{I\b} 
~+~
{\bar \si}^{~}_{I\m} \g^{~}_\n D^{~}_\a(\w(e)) \d \si^I_\b \cr 
&&~~~~~~~~~~~~~~~~~~~~~~+~ 
 \d{\bar \si}^I_\m \g^b_{~} \si^{~}_{I \b} ~\k^{~}_{\a\n b} ~+~ \d {\bar \si}^I_\m \si^J_\n~ {\bar \si}^{~}_{I\a} \si^{~}_{J\b} \left. \right]
\en

\noindent where the last two terms can be checked to cancel against 
each other by using the explicit expression for the $N=2$ contorsion 
 tensor (\ref{N2contorsion})
and a Fierz rearrangement. Again we notice that the presence of the
four-gravitino term in the $N=2$ super-Holst action (\ref{N2Haction})
is important for this cancellation to happen. 
Now the first two terms in the right hand side of above equation combine into 
a total derivative:
\bn
\left[ \d \si^I_\m~ {\frac {\d {\cal L}^{~}_{SHolst2}} {\d \si^I_\m}} \right]_{
\w=\w(e, \si)} ~=~ -~{\frac {i\eta} 2} ~ \e^{\m\n\a\b} ~
\partial^{~}_\m \left( \d {\bar \si}^I_\n \g^{~}_\a \si^{~}_{I\b}
\right)
\en
\noindent Hence this variation does not contribute to the gravitino
equation of motion; only contributions to the variation of total
action $S_2^{~}$ of Eqn.(\ref{N2action}) come from 
the  supergravity action $S^{~}_{SG2}$ (\ref{N2SGaction}) yielding 
the standard supergravity equations.

Similar conclusion holds for the other equation of motion obtained by 
varying the tetrad field $e^a_\m$. This can be seen explicitly from
\bn
\left[ \d e^a_\m~ {\frac \d {\d e^a_\m}}~ \left(e~\Sigma^{\m\n}_{ab}~ 
{\tilde R}^{ab}_{\m\n}(\w) \right) \right]_{\w =\w(e, \si)} 
=~ 2~ \e^{\m\n\a\b}_{~} \left[ \nabla_\m \k_{\a\b\l}^{~} + 
\k_{\m\b~}^{~~~\s}~\k^{~}_{\a\s\l} \right] ~e^\l_b \d e^b_\n
\en
\noindent and
\bn
-~\left[ \d e^a_\m~ {\frac \d {\d e^a_\m}}~\left( e^{\m\n\a\b} \left(
{\bar \si}^I_\m \g^{~}_\n D^{~}_\a(\w)\si^{~}_{I\b}~ +~ {\bar \si}^{~}_{I\m}
\g^{~}_\n D^{~}_\a(\w)\si^I_\b \right) \right)\right]_{\w=\w(e,\si)} \\
~~~~~=~ \e^{\m\n\a\b}~\left[ ~\nabla_\m \left({\bar \si}^I_\a \g^{~}_\l
\si^{~}_{I\b} \right) ~-~ {\bar \si}^I_\m \g^\s_{~} \si^{~}_{I\b} ~ 
\k^{~}_{\a\l\s} \right] ~e^\l_b \d e^b_\n
\en
\noindent From the expression for  contorsion tensor (\ref{N2contorsion}) notice that $ \e^{\m\n\a\b}_{~} 
\left({\bar \si}^I_\a \g^{~}_\l \si^{~}_{I\b} \right)$ $=~ -2~ 
\e^{\m\n\a\b}_{~}~\k^{~}_{\a\b\l}$ and $ e^{\m\n\a\b}_{~}
{\bar \si}^I_\m \g^\s_{~} \si^{~}_{I\b} ~ \k^{~}_{\a\l\s} $
$ =~ 2~ \e^{\m\n\a\b}_{~} \k_{\m\b}^{~~~\s}~ \k^{~}_{\a\s\l}$,
~ so that adding above two equations yields:
\bn
\left[ \d e^a_\m ~{\frac {\d {\cal L}^{~}_{SHolst2}} {\d e^a_\m}} \right]_{
\w=\w(e, \si)} =~ 0
\en
\noindent Again the $\d e^a_\m $ variation of total
action $S^{}_2$ obtains contributions only from the supergravity action 
(\ref{N2SGaction}) leading to the standard supergravity 
equation of motion.
Also, since the super-Holst action $S^{~}_{SHolst2} $ (\ref{N2Haction}) does 
not depend on the gauge field, the last equation of motion obtained by 
varying $A_\m$ comes from the supergravity action $S^{~}_{SG2}$
(\ref{N2SGaction}).

\section{N~=~4  supergravity}

Now we shall consider the generalisation of Holst action to the case
of $N=4$ supergravity \cite{N4SUGRA}. This theory, in its
$SU(4)$ version, describes four spin $3/2$ Majorana
gravitinos whose chiral projections
$\si^I_\m$ and $ \si^{~}_{I\m}$ ($I = 1, 2, 3, 4$) with $\g^{~}_5 \si^I_\m
= + \si^I_\m$ and $\g^{~}_5\si^{~}_{I\m} = - \si^{~}_{I\m}$ transform
as $4$ and ${\bar 4}$ representations of $SU(4)$ and four Majorana spin
$1/2$ fermions
whose chiral projections $\L^I_{~}$ and $\L^{~}_I$ with $\g^{~}_5\L ^I_{~} =-\L ^I_{~}$ and
$\g^{~}_5 \L^{~}_I = + \L^{~}_I$ also transform as $4$ and ${\bar 4}$
respectively. Bosonic fields of the theory include the tetrad fields $e^a_\m$
and six complex vector fields $A^{~}_{\m IJ}$ (antisymmetric in $IJ$)
and their $SU(4)$ dual
${\bar A}_\m^{~IJ} = \left( A^{~}_{\m IJ} \right)^*$ $ ={\frac 1 2}~
\e^{IJKL}_{~}~A^{~}_{\m KL}$. In addition, there are scalar fields that
parametrise the coset manifold $SU(1,~1) / U(1) $. These are represented as 
a doublet of $SU(1,~1)$  complex scalar 
fields $\p^{~}_A = (\p^{~}_1,~ \p^{~}_2)$ and their $SU(1,~1)$ dual
$\p^A_{~} = \eta^{AB}_{~} ~ \p^*_B$ $= (\p^*_1,~- \p^*_2)$ subject to the
condition $\p^A_{~}~\p^{~}_A $ $\equiv \p^*_1~\p^{~}_1 - \p^*_2~ \p^{~}_2$ $=1$.
The equations of motion of this theory exhibit an $SU(1, ~1)$ invariance, 
though its action does not. The action is given by \cite{N4SUGRA}:
\beq
S^{~}_{SG4} &=& \int d^4x ~e \left[ {\frac 1 4} ~R(\w, e) -
{\frac 1 {2e}}
\e^{\m\n\a\b}_{~} {\bar \si}^I_\m \g^{~}_\n {\cal D}_\a(\w) \si^{~}_{I\b} 
-{\frac 1 2}{\bar \L}^I_{~} \g^\m_{~} {\cal D}_\m^{~} (\w) \L^{~}_I
~~~~~~~~~~~~~~~~\right.\cr
&&~~~~~~~~~~~~~~~-~ {\frac 1 2}c^{~}_\m {\bar c}^\m_{~}
~-~{\frac 1 8} \left({\frac {\p^1_{~} -\p^2_{~}} {\P}}
\right)
F^{+}_{IJ\m\n}{\bar F}^{+IJ\m\n}_{~}  ~~~\cr
&& ~~~~~+ {\frac 1 {2 {\sqrt 2} 
\P}}{\bar \si}^I_\m \si^J_\n \left(
F^{+\m\n}_{IJ} + {\hat F}^{+\m\n}_{IJ} \right)  
-{\frac 1 {2 \P^*}} {\bar \L}^I_{~}\g_\m^{~} \si^J_\n
\left( F^{-\m\n}_{IJ} + {\hat F}^{-\m\n}_{IJ} \right) \cr
&& \left. -~{\frac 1 { \sqrt 2}}~{\bar \L}^I_{~} 
\g^\m_{~}\g^\n_{~} \left( c^{~}_\n + {\frac 1 {2 {\sqrt 2}}}{\bar \si}^J_\n 
\L^{~}_J \right) 
\si^{~}_{I\m} ~+~ c.~c. \right] 
\label{N4SGaction}
\eeq
\noindent where $\P \equiv (\p^1 +\p^2)$ and $\P^* \equiv (\p_1 -\p_2)$
and covariant derivatives ${\cal D}$ are:
\bn
{\cal D}_\m(\w) \L_I =\left(D_\m(\w) + (3i/ 2) a_\m \right)\L_I, ~~~~~
{\cal D}_\m(\w) \L^I = \left( D_\m(\w) - (3i / 2) a_\m \right) 
\L^I \\
{\cal D}_\a (\w) \si^I_\b = \left( D_\a (\w) + ( i/ 2) a_\a \right) 
\si^I_\b,
~~~~~ {\cal D}_\a (\w) \si^{~}_{I\b} = \left( D_\a (\w) - ( i/ 2) a_\a
\right) \si^{~}_{I\b}
\en

\noindent and the $SU(1,1)$ invariant vectors $ a^{~}_\m$, $c^{}_\m$ and
${\bar c}^{}_\m$ are:
\bn
a_\m ~=~ i \p^{~}_A \partial^{~}_\m\p^A_{~}, ~~~~~ c^{~}_\m ~=~ \e^{~}_{AB} 
~\p^A_{~} \partial^{~}_{\m} \p^B_{~}, ~~~~~ {\bar c}_\m^{~} ~=~ \e^{AB}_{~}
~\p^{~}_A \partial^{~}_\m \p^{~}_B
\en

\noindent The field strengths ~$F_{\m\n IJ}= \partial_{[\m} A_{\n] IJ}$
~and ~${\bar F}^{IJ}_{\m\n} = \partial_{[\m} A_{\n]}^{IJ} $ ~are 
super-covariantized as:
\bn
{\hat F}^{\m\n}_{IJ} ~=~ F^{\m\n}_{IJ} ~-~ {\frac 1 {2 {\sqrt 2}}}~ \P
\left( {\bar \si}^{[\m}_{[I} \si^{\n]}_{J]} + {\sqrt 2}~ \e^{~}_{IJKL}
{\bar \si}^{K[\m}_{~} \g^{\n]}_{~} \L^L_{~} \right) ~ \\
-~ {\frac 1{ 2{\sqrt 2}}} ~\P^* \left( \e^{~}_{IJKL} {\bar \si}^{K[\m}_{~}
\si^{\n] L}_{~} + {\sqrt 2}~ {\bar \si}^{[\m}_{[I} \g^{\n]}_{~} \L^{~}_{J]} \right) \\
{\hat {\bar F}}^{IJ}_{\m\n} ~=~ {\bar F}^{IJ}_{\m\n} ~-~ {\frac 1{ {2 {\sqrt 2}}}} 
~\P^* \left( {\bar \si}_{[\m}^{[I} \si_{\n]}^{J]} + {\sqrt 2} 
~\e_{~}^{IJKL}~ {\bar \si}_{K[\m}^{~} \g_{\n]}^{~} \L_L^{~} \right) ~ \\
		-~ {\frac 1{ 2{\sqrt 2}}} ~\P \left( \e_{~}^{IJKL}~ {\bar \si}_{K[\m}^{~}
				\si_{\n] L}^{~} + {\sqrt 2}~ {\bar \si}_{[\m}^{[I} \g_{\n]}^{~} \L_{~}^{J]} \right) 
		\en

To the $N=4$ supergravity action (\ref{N4SGaction}) we add 
a appropriately modified  Holst term:
\be
S_4~=~ S^{~}_{SG4}~+~ S^{~}_{SHolst4} \label{N4action}
\ee
\noindent where the $N=4$ super-Holst action is given by:
\beq
&&S^{~}_{SHolst4} ~=~ i\eta \int d^4x ~e~ \left[ {\frac 1 2}~ \Sigma^{\m\n}_{ab}
~{\tilde R}_{\m\n}^{~~~a b} (\w)~~~~~~~~~~~~~~~~~~~~~~~~~~ 
\right. \cr
&&~~~~~~~~~~~~~~~~~~~~~~~~~~~\left. - {\frac 1 {2e}}~ \e^{\m\n\a\b}_{~}
~\left({\bar \si}^I_\m \g^{~}_\n {\cal D}^{~}_\a (\w)\si^{~}_{I\b}
+ {\bar \si}^{~}_{I\m} \g^{~}_\n {\cal D}^{~}_\a (\w) \si^I_\b \right) 
\right.  \cr
&&~~~~~~~~~~~~~~~~~~~~~~~~~~~~-  {\frac  1 2}~ \left({\bar \L}^{~}_I \g^\m_{~} 
{\cal D}_\m^{~} (\w) \L^I_{~}
-{\bar \L}^I_{~} \g^\m_{~} {\cal D}_\m^{~} (\w) \L^{~}_I \right) \cr
&&~~~~~~~~~~~~~~~~~~~~~~~~~~~~  -{\frac 1 {4e}}~ \e^{\m\n\a\b}_{~} ~{\bar \si}^I_\m \si^J_\n 
~{\bar \si}^{~}_{I\a} \si^{~}_{J\b} ~\cr
&& ~~~~~~~~~~~~~~~~~~~~~~~~~~~\left. -~ {\frac 1 {4e}} ~\e^{\m\n\a\b}_{~}
~{\bar \L}^I \g_\m \si^J_\n ~{\bar \L}_I \g_\a \si_{J\b} \right]
\label{N4Haction}
\eeq
\noindent Here only those fields which are coupled to
 connection $\w$ in the supergravity action are involved
and not others like the gauge fields ~$A_{\m IJ}^{}$,
~ $ {\bar A}^{~IJ}_{\m}$~ and scalar fields ~$\phi^A_{}$~ which do not have
any coupling to $~\w$. Also in addition to the four-gravitino term, 
which is also present in the super-Holst action for $N=2$ 
supergravity, we have an additional four-fermion term involving 
two gravitinos and two $\L$'s. Both these
terms are important to achieve the desired result that equations
of motion of $N=4$ supergravity theory are not modified in presence of
this super-Holst term.

Variation of total action $S_4$ (\ref{N4action}) with respect 
to the connection $\w_\m^{~ab}$ leads to:
\bn
&&\int d^4x
\left[~ \e^{\m\n\a\b}_{~} ~\left(D_\b^{~} (\w) \Sigma^{cd}_{\m\n} 
- {\frac 1 2}~ {\bar \si}^I_\m  e^{[c}_\n\g^{d]}_{~} \si^{~}_{I\b} \right)
 \right. \cr
&&~~~~~~~~~~~~~~~~~~~~~~~~~ \left.-~e~{\bar \L}_I^{~} e^{\a [c} \g^{d]}_{~} \L^I_{~} \right]
\left({\frac 1 2}\e^{~}_{abcd} +i\eta \d^{~}_{ac} \d^{~}_{bd} \right)
\d \w_\a^{~ab} ~=~0
\en
\noindent This implies the standard torsion equation of $N=4$ supergravity:
\be
D_{[\m}^{}(\w) e^a_{\n ]} = 2 T_{\m\n}^{~~a} =2[T_{\m\n}^{~~a}(\si)
+T_{\m\n}^{~~a}(\L)] \equiv {\frac 1 2} {\bar \si}^I_{[\m} \g^a_{} \si^{}_{\n ]I}
+ {\frac 1 {2e}} e^{a \a}_{} \e_{\m\n\a\b}^{} 
{\bar \L}^{}_I \g^\b_{} \L^I_{} \label{N4torsion}
\ee
which is solved by
\be
\w_{\m ab}^{~} ~=~ \w^{}_{\m ab}(e, \si,\L) ~\equiv~\w_{\m  a b}^{~}(e) ~+~ \k_{\m ab}^{~} \label{N4spin}
\ee

\noindent where $\w^{~}_{\m ab}(e)$ is the standard pure gravitational
spin-connection given by (\ref{spine}) and $N=4$ contorsion 
tensor $\k$ has contributions
from both the gravitinos $\si$ and  fermions $\L$:
\beq
&&\k_{\m\a\b} ~=~ \k_{\m\a\b}(\si) ~ +~ \k_{\m\a\b} (\L) \cr
&&\k_{\m\a\b}(\si) ~=~ {\frac 1 4} \left[ {\bar \si}^I_\a \g^{~}_\m \si^{~}_{I\b}
~+~ {\bar \si}^I_\m \g^{~}_\a \si^{~}_{I\b} ~-~ {\bar \si}^I_\m
\g^{~}_\b \si^{~}_{I\a} ~ + ~ c.c. \right] ~~~~~~~\cr
&&\k_{\m\a\b}(\L) ~=~-~ {\frac 1 {4e}~}\e_{\m\a\b\s}^{~}
~{\bar \L}^{~}_I \g^\s_{~} \L^I_{~} 
~~~~ \label{N4contorsion}
\eeq

Like in the earlier cases of $N=1$ and $N=2$ supergravity, 
for the connection $\w = \w(e, \si, \L) =\w(e) + \k(\si, \L)$  
super-Holst Lagrangian density ${\cal L}_{SHolst4}^{}$ (\ref{N4Haction})
is a total derivative. To demonstrate that this is so, notice that:
\beq
&&-{\frac 1 2}\left[
~\e^{\m\n\a\b}~\left( {\bar \si}^I_\m \g^{~}_\n 
{\cal D}^{~}_\a (\w)\si^{~}_{I\b} + {\bar \si}^{~}_{I\m} 
\g^{~}_\n {\cal D}^{~}_\a(\w) \si_\b^I \right) \right. \cr
&& \left.~~~~~~~~ {\frac {} {}}+~ e ~\left( {\bar \L}^{~}_I \g^\m_{~}
{\cal D}_\m (\w)
\L^I_{~} - {\bar \L}^I_{~} \g^\m_{~}{\cal D}^{~}_\m (\w) \L^{~}_I \right) 
\right]_{\w=\w(e, \si, \L)} \cr
&&~~~~~~~~~~-~{\frac 1 4} \e^{\m\n\a\b} \left[ {\bar \si}^I_\m \si^J_\n ~
{\bar \si}^{~}_{I\a}\si^{~}_{J\b}
+ {\bar \L}^I_{~} \g_\m^{~} \si^J_\n ~
 {\bar \L}_I^{~} \g_\a^{~} \si^{~}_{J\b} \right]  ~~~~~~\cr
  && ~~~~= -~ {\frac 1 2}~ \left[ \partial_\m^{} \left(J^\m_{}(\si)
  + J^\m_{}(\L) \right) ~ + ~\e^{\m\n\a\b}_{} T_{\m\n a}^{} T_{\a\b}^{~~~a} 
  \right]
  \label{N4action1}
\eeq
\noindent where $J^\m_{}(\si) = \e^{\m\n\a\b}_{}~{\bar \si}^I_\n \g^{}_\a 
\si^{}_{I\b}$ and $J^\m_{}(\L)= e{\bar \L}^{}_I \g^\m_{} \L^I_{}$.
Here we have used $2 T_{\m\n}^{~~~\l}(\si) = -\k_{[\m\n]}^{~~~~\l}(\si)$,
~ $T_{\m\n\a}^{}(\L)=-\k_{\m\n\a}^{}(\L)$  and identities 
$ e {\bar \L}^{}_I \g^\a_{}\L^I_{} ~\k_{\m~~\a}^{~~\m}(\si)$
$= 2~\e^{\m\n\a\b}_{}T_{\m\n}^{~~~a}(\si) T^{}_{\a\b a}(\L)$, ~
$\e^{\m\n\a\b}~T_{\m\n a} (\L)
 T_{\a\b}^{~~~a}(\L)$ $=~0$
and the following relation obtained   by
 Fierz rearrangements:
\beq
&&-{\frac 1 2}~\e^{\m\n\a\b}~T_{\m\n a} T_{\a\b}^{~~~a} = 
-{\frac 1 2}~\e^{\m\n\a\b} \left[ T_{\m\n a}(\si) T_{\a \b}^{~~~a}
(\si) + 2~T_{\m\n a}(\si) T_{\a \b}^{~~~a}(\L) \right] \cr
&&~~~=~ {\frac 1 4} ~\e^{\m\n\a\b}~ \left[
{\bar \si}^I_\m \si^J_\n ~{\bar \si}^{~}_{I\a}\si^{~}_{J\b}  
+ {\bar \L}^I_{~} \g_\m^{~} \si^J_\n ~ 
 {\bar \L}_I^{~} \g_\a^{~} \si^{~}_{J\b} \right] \label{N4action2}
 \eeq
 \noindent Notice that the two four-fermion terms
 of the super-Holst action (\ref{N4Haction}) have played
 an important role in allowing us to write the equation (\ref{N4action1}).
 Now  substituting this equation  into the super-Holst action
 (\ref{N4Haction}), we find that super-Holst action for $\w =\w(e,\si,\L)$
 takes the same special form as in the earlier cases: 
\beq
S^{~}_{SHolst4} \left[ \w(e, \si, \L) \right]~ =~-~ {\frac {i\eta} 2}
\int d^4x \left[ I^{}_{NY}~+~
\partial_\m^{~}  J^\m_{~} (\si, \L) \right]
\label{N4nieh-yan}\eeq

\noindent where $J_\m^{}(\si, \L) \equiv J_\m^{} (\si) +J_\m^{}(\L)$.
It is important to note that this axial vector density $J_\m^{} (\si, \L)$
is not the conserved axial current of the $N=4$ theory;
in fact the conserved current density associated with the axial $U(1)$ 
invariance of the theory is ${\cal J}_\m^{} = J^{}_\m(\si) + 3~J^{}_\m(\L)$.

Now for the Nieh-Yan invariant density we use 
\bn
I^{}_{NY} =\e^{\m\n\a\b}
\partial_\m^{}T_{\n\a\b}^{} = \e^{\m\n\a\b} \partial_\m^{}[T_{\n\a\b}^{}
(\si) +T_{\n\a\b}^{}(\L)] \\
= -{\frac 1 2} ~\partial_\m^{} [J^\m_{}(\si)
+3 J^\m_{}(\L)]
\en
\noindent where we have used the facts:
$ 2\e^{\m\n\a\b} T^{}_{\n\a\b}(\si) =
- J^\m_{}(\si)$, ~ $2\e^{\m\n\a\b}_{} T_{\n\a\b}^{}(\L) = -3 J^\m_{}(\L)$.
This thus leads us to:
\beq
 S^{~}_{SHolst4} \left[  \w(e, \si, \L)\right]  = ~-~ {\frac {i\eta} 4 }
\int d^4x ~\partial^{}_\m \left[ J^\m_{}(\si) ~-~ J^\m_{}(\L) \right] \cr
= ~{\frac {i\eta} 2} 
 \int ~d^4x~\e^{\m\n\a\b}~ \partial_\m \left[ T_{\n\a\b}(\si)
 - {\frac 1 3} ~T_{\n\a\b}(\L)\right] 
 \eeq

Next to check explicitly that the other equations of motion are not changed
in this case too, consider for example, the  $\L_I$-variation
of the super-Holst  Lagrangian density ${\cal L}_{SHolst4}$ from
Eqn.(\ref{N4Haction}):
\bn
&& \d \L^{~}_I {\frac {\d {\cal L}_{SHolst4}} {\d \L^{~}_I}}
~=~ - {\frac {i\eta} 2} ~e~\left[ \left( {\frac {} {}}
\d{\bar \L}^{~}_I \g^\m_{~} 
{\cal  D}_\m^{~} (\w)\L^I_{~} - {\bar \L}^I_{~} \g^\m_{~}
{\cal D}^{~}_\m (\w) \d \L_I^{~} \right) \right. \cr
&& ~~~~~~~~~~~~~~~~~~~~~~~~~~~~~~~~~~~~~\left.  -
~ {\frac 1 2} ~\left( {\bar \si}^I_\m \g^\m_{~} \si^{~}_{I\n}
+ {\bar \si}^{~}_{I\m} \g^\m_{~} \si^I_\n \right) \d {\bar \L}^{~}_J
\g^\n_{~} \L^J_{~} \right]
\en
\noindent where, in writing the second term on the right hand side,
we have used the Fierz rearrangement 
\bn e ~\left( {\bar \si}^I_\m \g^\m_{~} \si^{~}_{I\n}
+ {\bar \si}^{~}_{I\m} \g^\m_{~} \si^I_\n \right)~ {\bar \L}^{~}_J
\g^\n_{~} \L^J_{~}~ = ~- ~\e^{\m\n\a\b} ~{\bar \L}^I_{~} \g_\m^{~} \si^J_\n
~{\bar \L}_I^{~} \g^{~}_\a \si^{~}_{J\b}
\en
\noindent Now substituting $\w=\w(e, \si, \L)$  from (\ref{N4spin})
we obtain:
\bn
&&\left[ \d \L^{~}_I {\frac {\d {\cal L}_{SHolst4}} {\d \L^{~}_I}}\right]_{\w
=\w(e,\si,\L)} = - {\frac {i\eta} 2}e \left[ {\frac {} {}} 
\d{\bar \L}^{~}_I \g^\m_{~}
{\cal  D}_\m^{~} (\w(e))\L^I_{~}   \right. 
  - {\bar \L}^I_{~} \g^\m_{~} 
{\cal D}^{~}_\m (\w(e)) \d \L_I^{~}  ~~~~~~~~~~~~~~~~~~~\cr
&& \left. ~~~~~~~~~~~~ ~~~~~~~~~~~~~~~~~~ +~ \d {\bar \L}^{~}_I 
\g^\n_{~} \L^I_{~} 
\left( \k_{\m~\n}^{~\m~} - {\frac 1 2} \left( {\bar \si}^I_\m \g^\m_{~} 
\si^{~}_{I\n}
+ {\bar \si}^{~}_{I\m} \g^\m_{~} \si^I_\n \right) \right) \right]
\en

\noindent Using  (\ref{N4contorsion}) for $N=4$ contorsion tensor,
last two terms cancel leaving the first two terms which 
combine into a total derivative:
\bn
\left[ \d \L^{~}_I {\frac {\d {\cal L}_{SHolst4}} {\d \L^{~}_I}}\right]_{
\w=\w(e, \si, \L)} = ~- ~{\frac {i\eta} 2}~ \partial^{}_\m \left( 
e\d{\bar \L}^{~}_I \g^\m_{~} \L^I_{~} \right)
\en

Similarly, variation of super-Holst Lagrangian density (\ref{N4Haction})
with respect to the gravitino $\si^I_\m$ and tetrad $e^a_\m$ fields are:
\bn
&&\left[ \d \si^I_\m~ {\frac {\d {\cal L}^{~}_{SHolst4}} {\d \si^I_\m}} \right]_{
\w=\w(e, \si, \L)} ~=~ -~{\frac {i\eta} 2} ~ \e^{\m\n\a\b} ~
\partial^{~}_\m \left( \d {\bar \si}^I_\n \g^{~}_\a \si^{~}_{I\b}
\right) \cr
&&\left[ \d e^a_\m ~{\frac {\d {\cal L}^{~}_{SHolst4}} {\d e^a_\m}} \right]_{
\w=\w(e,\si, \L)} =~ 0
\en

Thus clearly all the equations of motion  obtained by varying
the modified supergravity action $S^{}_4$ (\ref{N4action})
are the same as those obtained by varying the supergravity action
$S^{~}_{SG4}$ (\ref{N4SGaction}) alone;
addition of the super-Holst action 
$S^{~}_{SHolst4}$ (\ref{N4Haction}) does not change these 
classical equations of motion. These 
are indeed independent of the Immirzi parameter.

\section{Concluding remarks}

We have extended the Holst action for pure gravity with Immirzi
parameter as its associated coupling constant to the case of 
supergravity theories. This has been done in a manner that the 
equations of motion of supergravity theories are not changed by 
such modifications of the original Holst action. This ensures
that supersymmetry is preserved and Immirzi parameter does 
not play any role in the classical
equations of motion. This is  unlike the case studied by
Perez and Rovelli and also by Freidel, Minic and Takeuchi
\cite{perez-rovelli} where a spin $1/2$ fermion is minimally 
coupled to gravity
in presence of original Holst action without any modification.
In such a situation, the equations of motion do develop dependence 
on Immirzi parameter.

For each of $N=1, 2, 4$ supergravity theories we find that the modified
Holst Lagrangian density becomes  a total derivative when we use the
connection equation of motion $\w =\w(e, ...) = \w(e) + \k(...)$ 
where ellipsis indicates the various fermions which introduce torsion 
in the theory. This total derivative takes a special form analogous 
to the one described by Mercuri for the case of spin $1/2$ fermions
(\ref{mercuriF}). 
It is given in terms of  Nieh-Yan invariant density 
and divergence of an axial fermion  current density:
\be
S^{}_{Holst} [ \w = \w(e, ...)] ~=~ -~{\frac {i\eta} 2}
\int d^4x
\left[ ~I_{NY}^{} ~+~
\partial_\m^{~}  J^\m_{~}(...) ~\right] 
\label{nieh-yanG}
\ee
\noindent The Nieh-Yan topological  density is the divergence of 
pseudo-trace axial vector associated with torsion: $I_{NY}^{} ~=~ 
\partial_\m^{}~[\e^{\m\n\a\b}_{}~ T^{}_{\n\a\b}]$. 

It is important
to emphasise that the modified Holst action on its own does not
have this special form (\ref{nieh-yanG}) and reduces to this form only for the connection
that satisfies the connection equation of motion.

For arbitrary real values of Immirzi parameter $\eta^{-1}_{}$,
the Holst action allows
a canonical formulation of pure gravity \cite{holst,barros} in terms of
a real Ashtekar-Barbero $SU(2)$ connection. For modified
Holst action for the gravity coupled to a spin $1/2$ fermion,
this has also been done \cite{mercuri}. Extension of such a canonical
formulation to $N=1$ supergravity has been presented by Tsuda 
in \cite{N1sholst}.
In the same spirit, for the modified Holst actions (\ref{N2Haction})
and (\ref{N4Haction}) for $N=2$ and $N=4$ supergravity theories,
a similar generalised Hamiltonian formulation can be developed.
Care needs to taken in this analysis to fix the gauge after the
proper constraint analysis is performed \cite{sam}.
\vskip1.5cm

\noindent{\bf Acknowledgements}

\vspace{0.5cm}

Useful discussions with Naresh Dadhich, Ghanashyam Date and
 T.R. Govindarajan are gratefully acknowledged.

\end{document}